\documentclass[prb,showpacs,amsmath,amssymb,aps,epsfig, twocolumn]{revtex4}

\usepackage{graphicx}
\usepackage{epsfig}

\begin{document}
\draft
\title{Quantum dot size dependent influence of the substrate orientation on the electronic and optical properties of InAs/GaAs quantum dots}
\author{V. Mlinar}
\email{vladan.mlinar@ua.ac.be}
 \affiliation{Departement
Fysica,Universiteit Antwerpen, Groenenborgerlaan 171, B-2020
Antwerpen, Belgium}
\author{F. M. Peeters}
\email{francois.peeters@ua.ac.be}
\affiliation{Departement
Fysica,Universiteit Antwerpen, Groenenborgerlaan 171, B-2020
Antwerpen, Belgium}

\date{\today}

\begin{abstract}
Using 3D \textbf{k.p} calculation including strain and piezoelectricity we predict variation of electronic and optical properties of InAs/GaAs quantum dots (QDs) with the substrate orientation. The QD transition energies are obtained for high index substrates [11k], where k = 1,2,3 and are compared with [001]. We find that the QD size in the growth direction determines the degree of influence of the substrate orientation: the flatter the dots, the larger the difference from the reference [001] case.
\end{abstract}


\maketitle

Semiconductor quantum dots (QDs) unique features make them
promising candidates for novel semiconductor
devices.~\cite{Shchukin} In the widely used Stranski- Krastanow
growth mode, growth conditions determine the electronic and
optical properties of QDs but also introduce size, shape, or
chemical composition uncertainty.~\cite{Shchukin1} So far, most
experimental~\cite{Fry} and
theoretical~\cite{Zunger,Stier,Leburton,Mlinar} studies were
performed on QDs grown on [001] substrates.

Recently, interest has moved towards QDs grown on high index
surfaces where a substantial amount of
experimental work has been done.~\cite{Gong,Godefroo,Jacobi1,Jacobi2}
Growth of QDs on high index planes has several practical
advantages. For example, growth on a [113]B substrate leads to
good quality QD structures with high densities and low size
dispersion which are useful for QDs based lasers.~\cite{Caroff} Very recently
it was found that planar and vertical ordering in QD lattices can
be controlled by substrate orientation enabling 3D growth ranging
from a chainlike pattern to a square-like lattice of
QDs.~\cite{Schmidbauer} From a physics point of view, different
substrate orientations result in different planar projections of
conduction and valence bands of the constituent crystals forming
QD's. As a consequence, the photoluminescence energy is expected
to change with the substrate orientation. It is of fundamental
importance to understand the underlying physical features of such
systems. How does the strain distribution in and around QDs depend
on substrate orientations? How are the electronic structure and
transition energies influenced by the substrate orientation? The
aim of this Letter is to answer these questions and to point out
the main differences with the well investigated [001]
grown QDs.

The influence of the substrate orientation is more pronounced if
the degree of lattice mismatch between the dot and the barrier is
larger as it is the case for InAs/GaAs QDs, where the lattice
mismatch is $ \sim7.2 \%$. We consider such InAs/GaAs QDs grown on
[11k] substrates, where k=1,2,3. 
We tested various dot shapes and sizes. Here we present the
results for two different dot shapes: lens and truncated pyramid,
and three different sizes: For lens shaped QDs: R = 6.78nm, h
=2.83nm (L1), R = 9.9nm, h=3.84nm (L2), R=10.17nm, h=10.17nm (L3),
and for truncated pyramid b=14.7nm, h=3.4nm (P1), b=18nm, h =
3.6nm (P2), b=22nm, h=4.5nm (P3).

\begin{figure}
\begin{center}
\epsfig{file=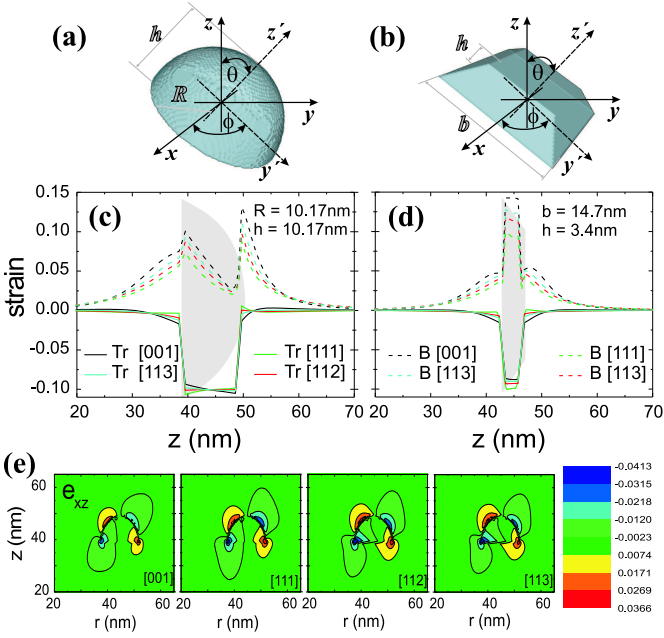,width=8 cm} \caption{(color online) For lens
shaped (a) and truncated-pyramidal (b) QDs, relationship of the
general [11k] coordinate system to the conventional [001]
coordinate system [$\phi = \pi/4$, $\theta=arctan(\sqrt{2}/k)$].
Isotropic Tr (solid lines) and biaxial part B (dashed lines) of
the strain tensor for lens-shaped (c) and truncated pyramidal (d)
QD. The gray surfaces in (c) and (d) represent the dots in the
growth direction. (e) e$_{xz}$ strain component of L3 QD in the
plane demonstrating an increase of the shear strain with changing
the substrate orientation.}\label{fig.1}
\end{center}
\end{figure}

In our 3D model, the strain distribution of the InAs/GaAs QDs is
calculated using continuum elasticity and single particle states
are obtained from eight-band \textbf{k.p} theory~\cite{Mlinar1}
including strain and piezoelectricity. In order to properly take
into account the effect of the different substrate orientation,
the coordinate system is rotated in a way that the Cartesian
coordinate z$'$ coincides with the growth direction
[Figs.~\ref{fig.1} (a) and (b)].~\cite{Henderson} The general [11k] coordinate system
$(x',y',z')$ is related to the conventional [001] system $(x, y, z)$ through a transformation matrix U=U$(\phi,\theta)$. The angles $\phi$
and $\theta$ represent the azimuthal and polar angles,
respectively, of the [11k] direction relative to the [001]
coordinate system.

{\it Strain distribution.} In order to determine the character of
the strain for dots grown on [11k] substrates, we decompose the
calculated strain tensor into an isotropic part
$Tr(e)=e_{xx}+e_{yy}+e_{zz}$ and a biaxial part
$B=\sqrt{(e_{xx}-e_{yy})^2+(e_{yy}-e_{zz})^2+(e_{zz}-e_{xx})^2}$,
where $e_{\alpha\beta}$ denotes the $\alpha\beta$ component of the
strain tensor. The strain profiles along the growth direction
across the lens shaped L3 and truncated pyramidal P1 QDs are shown
in Figs.~\ref{fig.1}(c) and (d) respectively. First, for [001]
grown QDs, the isotropic strain is negative (compressive) inside
the dot and tends to zero rapidly in the
barrier. 
This isotropic strain is increased in [11k] grown flat QDs
regardless of the dot shape, and the largest increase was found
for [111] grown dots. However, this is no longer the case for
larger dots [pyramidal, half-spherical, or conusoidal QDs], where
variation of the substrate orientation keeps the isotropic strain
almost constant in the growth direction of the dot. Second, for
[001] grown QDs the biaxial strain is non zero inside the dot and
reaches a significant amount into the barrier decaying very slowly
to zero. For flat dots the biaxial strain is almost constant
inside the dot [Fig.~\ref{fig.1}(d)], while for the larger dots it
has a distinct minimum in the QD [Fig.~\ref{fig.1}(c)]. QD growth
on [11k] surfaces does not modify such a behavior of biaxial
strain but just decreases the biaxial component regardless of the
dot size and shape.
Third, for all the considered dot sizes and shapes shear strains
are increased for [11k] growth. As an example, we show in
Fig.~\ref{fig.1}(e), for the L3 QD, the e$_{xz}$ strain component
as it varies with substrate orientation. These shear components
leads to a strongly asymmetric piezoelectric potential for [11k]
grown QDs. What are the consequences of the different strain
distributions in [11k] grown QDs as compared to [001] grown QDs?
The isotropic part of the strain tensor shifts the conduction band
upwards and the valence band downwards. Therefore, one can expect
that only the electron states of [11k] grown flat dots will lie
energetically higher  as compared to the [001] grown QDs. Biaxial
strain determines the heavy hole - light hole band splitting in
the bulk, but for [11k] substrate orientation the different
valence bands interact even at the zone center (see below),
therefore, the lower the biaxial strain, the larger the mixing of the different valence bands.
Furthermore, the asymmetric piezoelectric potential influences the
distribution of the electron and hole wavefunction inside the dot.

\begin{figure}
\begin{center}
\epsfig{file=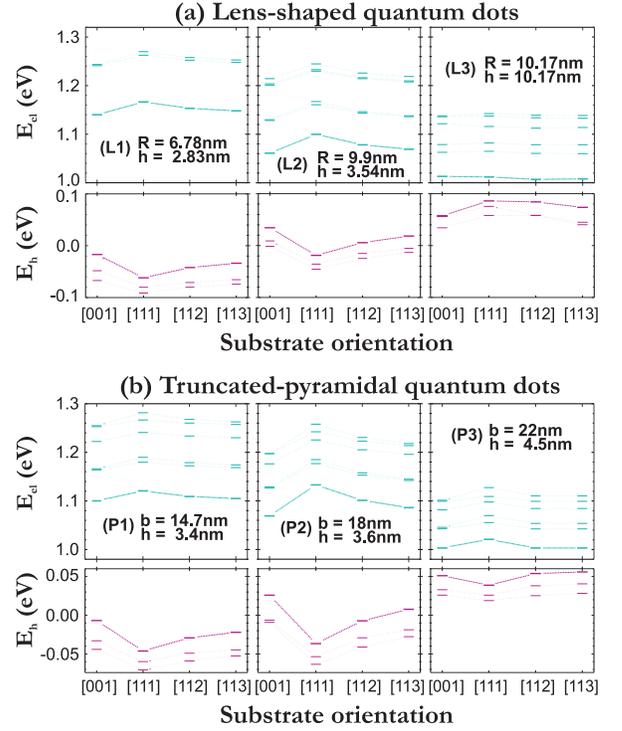, width=8 cm} \caption{(color online)
Electron and hole energy levels as they vary with the substrate
orientation for lens shaped (a) and truncated pyramidal (b) QDs.
Electron and hole energies are given with respect to the top of
the valence band of InAs.}\label{fig.2}
\end{center}
\end{figure}

{\it Electronic structure.} Using a diagonalization of the
eight-band Hamiltonian, including the strain and the piezoelectric
potential, confined electron and hole energy levels are obtained
numerically, which are shown in Fig.~\ref{fig.2}.
In the upper panels of Figs.~\ref{fig.2}(a) and (b), the lowest
lying electron energy levels of lens shaped and
truncated-pyramidal QDs are shown. Variation of the electron
energy levels with the substrate orientation depends on the dot
size in the growth direction. For smaller dots the influence of
the substrate orientation on the electron energy levels is larger.
Note that this is consistent with the analysis of the strain
distribution in and around the QD as a function of the substrate
orientation and QD size and shape. The situation with the hole
states is more complex. In the lower panels of
Figs.~\ref{fig.2}(a) and (b) the hole energy levels are shown. It
is important to stress that, for [11k] substrate orientation,
different valence bands interact even at the zone center, implying
that the hole states at the zone center can not be classified as
pure heavy or pure light hole as in the case for [001]
substrate orientation. Compared to [001] growth, there is
increased valence band mixing induced by the kinetic part of the
Hamiltonian and due to the reduced biaxial component of the strain, and
the increased isotropic part of the strain tensor. Similar to the case for
the electron states, the QD size determines the variation of the
hole energies with substrate orientation. The flatter the dots
are, the larger the difference with the hole energy levels of
the reference [001] case. What are the consequences of such
changes for the electronic structure? It is expected that the
transition energies of a [11k] grown flat dot becomes
significantly different from the transition energies of [001]
grown dots. Larger QDs do not show such a behavior and their
transition energies are expected to be closer to the one of the
[001] grown QD.

\begin{figure}
\begin{center}
\epsfig{file=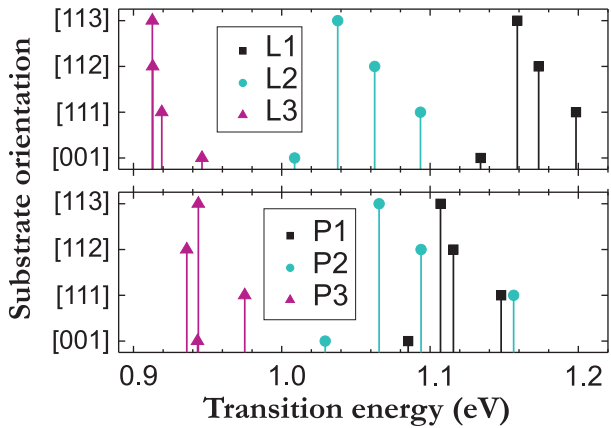,width=8 cm} \caption{(color online)
Transition energies as they vary with the substrate orientation
for lens shaped (upper panel) and truncated pyramidal (lower panel) QDs.}\label{fig.3}
\end{center}
\end{figure}

{\it Transition energies.} The dependence of the single particle
energy levels on the substrate orientation is different for
different QD sizes. Therefore, we expect that in an ensemble
of QDs the emission from the larger dots is more dominant.
Including direct Coulomb interaction in our calculations,
the variation of the transition energy with the substrate
orientation and QDs size and shape is shown in Fig.~\ref{fig.3}.
For flat dots, the largest difference between the transition
energies of [11k] grown QDs and with the reference [001] case is
found for the [111] grown QD, whereas [113] grown QDs have
transition energies that are closest to the ones of [001]. This
conclusion is no longer valid when the size of the dots in the
growth direction is increased. One can see from Fig.~\ref{fig.3}
that the transition energies of [11k] grown L3 QDs, are close to
each other, and lower as compared to the transition energy of
[001] grown QDs. A similar situation occurs for the transition
energies of P3 QDs, where the transition energies of [112] and
[113] grown QDs are lower than the one of the [001] grown QD, but
the transition energy of the [111] grown QD is higher. This is a
consequence of the lower height of the P3 QD as compared to the L3 QD.

In conclusion, our 3D \textbf{k.p} calculation including strain and
piezoelectricity predict the dependence of the transition energies of
InAs/GaAs QDs on substrate orientation. We show that the QD size in the
growth direction determines the degree of the influence of the
substrate orientation on the electronic and optical properties of
[11k] grown QDs, whereas the influence of the shape is of secondary
importance. The flatter the dots are, the larger the difference from the reference [001] case.
Although the composition intermixing and shape
variation related to the growth conditions can quantitatively
influence our results, the presented work should be understood as
a guideline for the variation of the electronic and optical
properties of QDs going from well investigated [001] grown QDs to
[11k] grown QDs, where k = 1,2,3.


This work was supported by the Belgian Science Policy (IUAP), and
the European Union Network of Excellence: SANDiE.

\end{document}